\documentclass[conference]{IEEEtran}
\usepackage{etex}
\usepackage{tikz}
 \usepackage[T1]{fontenc}
\usepackage{tikz-qtree}
\usepackage{epigraph}
\usepackage{adjustbox}
\usepackage{setspace}
\usepackage{cite}
\usepackage[latin1]{inputenc}
\usepackage{epsf}\epsfverbosetrue
\usepackage{graphics,epsfig,color}
\usepackage{graphicx,epsfig}
\usepackage{epsfig}
\usepackage{multirow}
\usepackage{slashbox}
\usepackage{alltt}
\usepackage{subfigure}
\usepackage{color}
 \usepackage{float} 
\usepackage{url}
\usepackage{graphicx}
\usepackage{verbatim} 
\usepackage{footnote} 
\usepackage{latexsym} 
\usepackage{amsmath}
\usepackage{sidecap} 
\usepackage{color}
\usepackage{wrapfig}
\usepackage{algorithm}
\usepackage{eso-pic}
\usepackage{fix-cm}
\usepackage{mcite}
\usepackage{amsfonts}
\usepackage{color}
\usepackage{wrapfig}
\usepackage{algorithm}
\usepackage{layout}
\usepackage{amsmath}
\usepackage{textcomp}
\usepackage{multirow}
\usepackage{soul,xcolor}
\usepackage{amsmath}
\usepackage{verbatim}
\usepackage{smartdiagram}
\usepackage{metalogo}
\usepackage{bbding}  

\usepackage{tikz}
\usetikzlibrary{shapes}
\usepackage{amsmath}
\usepackage{xspace}

\usepackage{tikz}
\makeatletter 
\@namedef{color@1}{red!40}
\@namedef{color@2}{green!40}   
\@namedef{color@3}{blue!40} 
\@namedef{color@4}{cyan!40}  
\@namedef{color@5}{magenta!40} 
\@namedef{color@6}{yellow!40}    

\newcommand{\graphitemize}[2]{%
\begin{tikzpicture}[every node/.style={align=center}]  
  \node[minimum size=5cm,circle,fill=gray!40,font=\Large,outer sep=1cm,inner sep=.5cm](ce){#1};  
\foreach \gritem [count=\xi] in {#2}
{\global\let\maxgritem\xi}  
\foreach \gritem [count=\xi] in {#2}
{%
\pgfmathtruncatemacro{\angle}{360/\maxgritem*\xi}
\edef\col{\@nameuse{color@\xi}}
\node[circle,
     ultra thick,
     draw=white,
     fill opacity=.5,
     fill=\col,        
     minimum size=3cm] at (ce.\angle) {\gritem };}%
\end{tikzpicture}  
}%

\newcommand*{\rom}[1]{\expandafter\@slowromancap\romannumeral #1@}
\newcommand{\Rmnum}[1]{\expandafter\@slowromancap\romannumeral #1@}
\makeatother

\begin{document}
\title{Performance Analysis of NOMA for Ultra-Reliable and Low-Latency Communications}
\author{\IEEEauthorblockN{Muhammad Amjad and Leila Musavian}
\IEEEauthorblockA{School of Computer Science
and Electronic Engineering (CSEE)\\ University of Essex, UK.\\
Email: \{m.amjad, leila.musavian\}@essex.ac.uk  }}
  \maketitle

{\let\thefootnote\relax\footnote{{Part of this work has been supported by H2020-MSCA-RISE-2018 project ``RECENT''.}}}
%

\begin{abstract}
Grant-free non-orthogonal multiple access (NOMA) has been regarded as a key-enabler technology  for ultra-reliable and low-latency communications (URLLC). In this paper, we analyse the performance of  NOMA with short packet communications for URLLC. In this regard, the overall packet loss probability consists of transmission error probability and queueing-delay violation probability.   Queueing-delay has been modelled using the effective bandwidth. Due to short transmission time, the infinite block-length has been replaced with finite blocklength of the channel codes which rules out the application of Shannon's formula. The achievable effective bandwidth of the system is derived, and then, the transmission error probability has been analysed. The  derivations are validated through extensive simulations, which shows the variations of the signal-to-noise ratio (SNR) requirements of the system for various transmission-error probability, QoS exponent, and the transmission  packet size. 
\end{abstract}

\IEEEpeerreviewmaketitle


\section{Introduction}
\label{sec:in}
As compared to the previous cellular system generations, fifth-generations (5G)  is considered to be more innovative while promising the seamless connectivity for massive machine type communications (mMTC) and  ultra-reliable and low latency communications (URLLC) \cite{R016}.  Considering URLLC,  achieving a low latency in conjunction with a very high reliability is a challenging task. However,   with latency as much low as 1ms and   99.999\% reliability, mission critical communications can be ensured. This feature can support the ballooning demands of critical applications such as Tactile Internet, autonomous vehicles, and factory automation \cite{R007}.

 Non-orthogonal multiple access (NOMA) technology has been introduced to address the need for high data rate while supporting multiple users within same time/frequency resource block \cite{R145}. When compared to the orthogonal multiple access (OMA), NOMA can provide higher performance in terms of spectral efficiency and energy efficiency. The potential benefits of NOMA can be exploited while integrated it with other   technologies such as short-packet communications. On the other hand,  grant-free NOMA has been proposed as a potential enabler technology for URLLC to support the time-critical applications \cite{R136}.

Researchers have come out with different designs paradigm to achieve the URLLC. This technology, infact, is envisioned as a potential solution for achieving low-latency in URLLC. With the short packet communications of finite blocklength, the conventional shannon capacity has been replaced with a new formula proposed in \cite{R139}.

In most of the existing studies, only the transmission delay and transmission error probability have been taken into account to model the required delay and reliability for the URLLC. However,  we note that the queueing-delay  could consist the highest latency in the system.   In this regard, the concept of  effective-bandwidth has been used to model the queuing delay of short packet communications in \cite{R007} which also considers the transmission error probability, queueing-delay violation probability, and   proactive packet dropping probability. The proposed scheme proactively drops some packets to ensure the high reliability and low latency.  Other designs paradigm has also been adopted to improve the transmission error probability in \cite{R141,R142,R143,R144}. Particularly,  the polar codes are introduced in \cite{R141}  to minimize the various delays that caused from the signal processing, coding, and transmission. Various diversity schemes  have also been exploited to minimize the transmission and other delays while supporting the URLLC. Spatial diversity  has been used to improve the reliability in \cite{R142}, while the frequency diversity has been investigated in \cite{R143}. Further, \cite{R144}  shows that delay and reliability requirements  for URLLC can be achieved with increased number of antennas at the BS. 

On the other hand,  performance of NOMA with short packet commications has been investigated in \cite{R140} where in the Transmission error probability has been translated into the block error length with finite blocklength and short transmission time. Simulations results demonstrate the effectiveness of the NOMA  as compared to the OMA for reducing the transmission latency. However, the queueing-delay and queueing delay violation probability has not been taken into consideration in modelling the reliability and latency requirements in \cite{R140}. 

With the above background study coverage, in this paper, we aim to answer the following question:

\textit{Given the low latency and high reliability requirements of URLLC, how much latency can NOMA reduce while satisfying the given reliability constraints. }

Employing the NOMA with short packet commications can result in the following advantages,

\begin{itemize}
 \item As compared to  NOMA, in conventional OMA schemes, base station (BS) has to employ the contention-based access scheme to mitigate the collision among multiple users. This can result into the increased delay/latency due to the increased collisions with the increased number of users. Hence, for the ultra-low latency, OMA cannot be employed.
 
 \item NOMA does not use the grant acquisition as it uses the grant-free access. This grant-free access is helpful in satisfying the required delay and reliability requirements when the number of users increases. 
 
 \item NOMA can also be easily integrated with the other coding schemes such as polar codes to achieve the low latency for ultra-delay sensitive applications. 
\end{itemize}
This paper will be a first attempt to analyse the performance limitations of NOMA in short packet communications for URLLC. The major contributions of our work can be summarized as follows:

\begin{itemize}

 \item A performance analysis of NOMA for URLLC has been provided while considering the  queueing-delay, error-rate, and packet sizes.  
 \item Transmission error probability and queueing-delay violation probability with finite blocklength codes has been taken into consideration to ensure the overall reliability. 
 \item  Analysis of effective-bandwidth for the NOMA in URLLC has been provided. This analysis is then used to find the required SNR for the given latency and erro-rate. 
 \item Extensive simulations have been carried out to investigate the performance of NOMA for URLLC while taking into consideration the required SNR, transmission-error probability, delay exponent, and different packet sizes. 
 
\end{itemize}
The rest of the paper is organized as follows. Section \ref{sec:system-model} provides the system model and problem formulation. Section \ref{sec:per-evaluation} shows the performance evaluation of the proposed analysis, and Section \ref{sec:conclusion} concludes the paper with future research directions.  
\section{System Model and Problem Formulation}
\label{sec:system-model}
We consider a two users  power-domain NOMA operation.  Out of the two users, one user is getting a better channel condition such that,
\begin{equation}
 \left | h_{1}  \right |^2\geq \left | h_{2}   \right |^2,
\end{equation}
where $h_{i}$ is the channel coefficient between the user $k_i$  and the BS, where $i=\left \{ 1,2 \right \}$. 
According to the NOMA operation, the broadcast signal from the BS to the users can be defined as: 
\begin{equation}
 x=\sum_{i=1}^{2}\sqrt{\alpha_{i}Pm_{i}},
\end{equation}
where $\alpha_i$ is the power allocation coefficient, $P$ is the transmitted power, and $m_i$ is the message of user $k_i$. The received signal at user $i$ is hence given by
\begin{equation}
y_i=h_i\sum_{i=1}^{2} \sqrt{\alpha_{i}Pm_i} + n_i,
\end{equation}
where $n_i$ is the additive white Gaussian noise at user $i$. According to the conventional NOMA operation, $k_2$ decodes the $m_2$ first. The resulting  signal-to-interference-plus-noise ratio (SINR)  at $k_2$ to decode $m_2$, hence  can be approximated as follows, 
 \begin{equation}
\label{eq:snr22}
 \gamma_{22}=\frac{\alpha_{2}\left | h_{2} \right |^2 }{(\alpha_{1}\left | h_{2} \right |^2+\frac{1}{\rho })} 
\end{equation}
where $\rho$ is the  transmit  signal-to-noise ratio (SNR). Similarly,   $k_1$  first decodes $m_2$. The received SINR at $k_1$ for decoding first the $m_2$ is given as follows, 
\begin{equation}
\label{eq:snr12}
 \gamma_{12}=\frac{\alpha_{2}\left | h_{1} \right |^2 }{(\alpha_{1}\left | h_{2} \right |^2+\frac{1}{\rho })},
\end{equation}
after successfully decoding the $m_2$, the received SNR at $k_1$ to decode  $m_1$ is, 
\begin{equation}
\label{eq:snr11}
 \gamma_{11}=\alpha_{1}\rho\left | h_1 \right |^2 .
\end{equation}
\noindent The overall reliability is the overall packet loss  probability $\varepsilon _{D}$ of a single user which is the combination of transmission error probability and queueing-delay violation probability:   
\begin{equation}
\label{eq:overallreb}
\varepsilon _{D}=\varepsilon _{C}+\varepsilon _{Q},
\end{equation}
where $\varepsilon _{C}$ is the transmission-error probability and $\varepsilon _{Q}$ is the queuing-delay violation probability.  Normal operation of NOMA revolves around the principle of superposition coding (SC) at the transmitter and interference cancellation at the receiver. After considering the queueing delays and successive interference cancellation at the two users, the overall reliability can be summarized as follows,  
\begin{equation}
 \varepsilon _D=\varepsilon _{C_{ij}}+\varepsilon _{Q_{ij}},
\end{equation}
where  $\varepsilon _{C_{ij}}$ and $\varepsilon _{Q_{ij}}$  are the transmission error probabilities and queuing-delays violation probabilities of user $k_i$ to decode the message $m_j$, where $i,j=\left \{ 1,2 \right \}$.
\noindent In this paper, we use the concept of effective bandwidth which models the performance of the system when taking into consideration the queuing-delay violation probability. Effective bandwidth can be defined as the  minimal constant service rate while satisfying a certain queueing delay constraint for a given arrival process. It can be derived through large-deviation approximation. For poisson arrival process \{$ A_{p}$\}, the effective-bandwidth can be defined as \cite{R146}  
\begin{equation}
\begin{split}
E_{ij}^{B}(\theta_{ij})=\frac{1}{T_{f} \theta_{ij}} {\rm ln}  \left \{ \mathbb{E} \left [ {\rm exp}\left (  \theta_{ij}  A_{p}\right ) \right ]   \right \},
\end{split}
\end{equation}
where $\theta_{ij}$ is the quality-of-service (QoS) exponent for the user $k_i$  to decode the message $m_j$, smaller value of $\theta_{ij}$ indicates the larger queuing-delay bound. $T_{\rm{f}}$ is the frame duration and  $\mathbb{E}[.]$ is the expectation operator.  If the arrival rate is constant then the queueing-delay violation probability can be derived as \cite{R147}
\begin{equation}
\label{eq:qdviolation}
 {\rm Pr}\left \{ D_{ij}(\infty)> D_{{\rm max}}^{q} \right \}\approx \eta_{ij}{\rm exp}\left \{-\theta_{ij} E_{ij}^{B}(\theta_{ij})D_{{\rm max}}^{q}  \right \},
\end{equation}
where Pr$\{a>b\}$ shows the probability of $a$ being bigger than $b$, $D_{{\rm max}}^{q}$ is the delay bound, $\approx$ shows the approximation,  and  $\eta_{ij}$ is the probability of non-empty buffer and is approximated accurately  when the queue length tends to infinity.  For $\eta_{ij}\leq 1$, the queueing-delay violation probability can be, 
\begin{equation}
  {\rm Pr}\left \{ D_{ij}(\infty)> D_{{\rm max}}^{q} \right \}\approx{\rm exp}\left \{-\theta_{ij} E_{ij}^{B}(\theta_{ij})D_{{\rm max}}^{q}  \right \}. 
\end{equation}
From (\ref{eq:qdviolation}), the queuing-delay violation probability for user $k_i$ to decode message $m_j$ can be approximated as,  
\begin{equation}
 \varepsilon _{Q_{ij}}\approx{\rm exp}\left \{-\theta_{ij} E_{ij}^{B}(\theta_{ij})D_{{\rm max}}^{q}  \right \}.
\end{equation}
The maximum number of packets that can be transmitted to any single user in  frame $n$ can be approximated as \cite{R118} 
\begin{equation}
\begin{split}
s_{ij}(n)=\frac{\phi B}{u{\rm ln}(2)} 
\left \{{\rm ln}\left [ 1+\frac{\alpha_{i} P h_{i}}{N_{o} B} \right ]  -  \sqrt{\frac{V_{ij}}{\phi B}}Q^{-1}{(\varepsilon_{C_{ij}})}\right \} \\ {\rm (packets)},
\end{split}
\end{equation}
\noindent   where $\phi$ is the data transmission phase,  $B$ is the bandwidth. $\phi B$ is the block-length that can be used for the short packet communications, $u$ is the size of packet in bits,  $N_{o}$ is the spectral density of noise, $Q^{-1}(.)$ is the inverse of Gaussian Q-function,  and $V_{ij}$ is the channel dispersion of user $k_i$ to decode the message $m_j$, that can be approximated by  \cite{R148}
\begin{equation}
 V_{ij}=1-\frac{1}{[ 1+\frac{\alpha_{i} Ph_{i}}{N_{o} B}]}.
\end{equation}
If the $k_{th}$ user is served with the constant rate, i.e., $E_{ij}^{B}(\theta_{ij})$, then we can approximate the number of packets transmitted in  frame $n$ using \cite{R007}, 
\begin{equation}
\label{eq:arrival-approximate}
 s_{ij}(n)=T_{f}E_{ij}^{B}(\theta_{ij}) \quad {\rm (packets)}.
\end{equation}
From (\ref{eq:snr22}), (\ref{eq:snr12}),  (\ref{eq:snr11}), and (\ref{eq:arrival-approximate}), the required SINR at user $k_i$ to decode message $m_j$  can be approximated using:
\begin{equation}
\begin{split}
 {\rm ln}(1+\gamma_{ij})=\frac{T_{f}u{\rm ln}(2)}{\phi B}E_{ij}^{B}(\theta_{ij})+ 
 \sqrt \frac{V_{ij}}{\phi B}Q^{-1}(\varepsilon_{C_{ij}}).
\end{split}
 \end{equation}
The performance of the proposed NOMA for achieving URLLC can be compared with the OMA for URLLC. For any single user $k_i$, the performance of orthogonal frequency-division multiplexing (OFDM) for URLLC can be derived  as:
\begin{equation}
\begin{split}
 s_i(n)=\frac{\phi B}{2u{\rm ln}(2)}  \left \{{\rm ln}\left [ 1+\frac{\alpha_{i} P h_{i}}{N_{o} B} \right ]  -\sqrt{\frac{2V_{i}}{\phi B}}Q^{-1}{(\varepsilon_{C_{i}})}\right \},
 \end{split}
 \end{equation}
 where $V_i$ and $\varepsilon_{C_{i}}$  is the channel dispersion and the transmission error probability  of the $k_{ith}$ user employing the OMA operation.  
From  (\ref{eq:arrival-approximate}) the required SNR to calculate the transmission error probability $(\varepsilon_{C_i})$ and  queueing-delay violation probability $(\varepsilon_{Q_i})$ for $k_{ith}$ user under OFDMA operation can be described as follows:
\begin{equation}
\begin{split}
 {\rm ln}(1+\gamma_{i})=\frac{2T_{f}u{\rm ln}(2)}{\phi B}E_{i}^{B}(\theta_{i})  +  \sqrt \frac{2V_{i}}{\phi B}fQ^{-1}(\varepsilon_{C_{i}}),
\end{split}
 \end{equation}
where $\gamma_{i}$ is SNR and $E_{i}^{B}(\theta_{i})$ is the effective-bandwidth  with $\theta_i$ as the QoS exponent of the $k_{ith}$ user.  
 
\section{Performance Evaluation}
\label{sec:per-evaluation}
\begin{table}
\footnotesize
\centering
\caption{Simulation Parameters}
\label{tab_sim_para}
\begin{tabular}{|p{3.7cm}|p{2cm}|}
\hline

 Frame duration  $T_f$    &	 0.5 ms     \\\hline
      
     Overall reliability $\varepsilon _{D}$       &	 1 - 99.999\%     \\\hline
       Queuing Delay $D_{\rm max}$           &	0.8 ms      \\\hline
          Bandwidth $B$              &100 kb/s	      \\\hline
           Packet size $u$                   &5,15, and 25 bytes      \\\hline
            Duration of data transmission $\phi$                        &	$10^{-5}$ and $3\times10^{-4}$      \\\hline
            
            Average number of events per event $\lambda$    &	0.01     \\\hline
      
             \end{tabular}
    \end{table}

\begin{figure}  
\centering
\includegraphics[width=8.5cm,height=7.5cm]{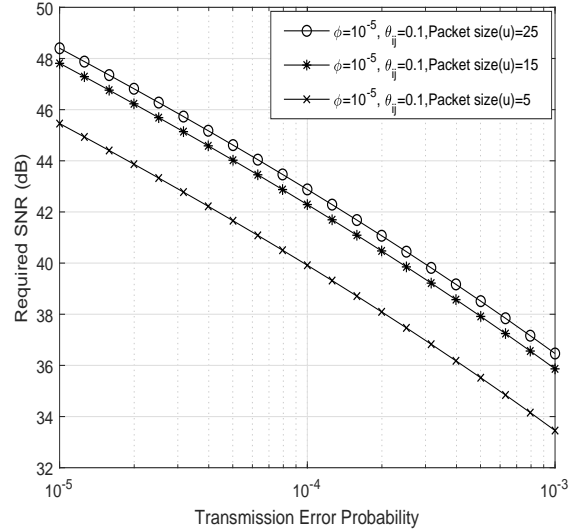}
\caption{SNR requirements for the given transmission error probability ($10^{-5}$ to $10^{-3}$)  with different  packet sizes. }
\label{Figure1}
\end{figure}
\begin{figure}  
\centering
  \includegraphics[width=8.5cm,height=7.5cm]{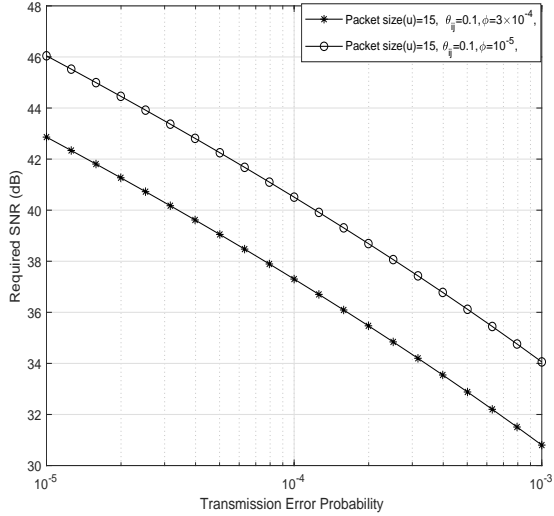}
\caption{SNR requirements for the given transmission error probability ($10^{-5}$ to $10^{-3}$)  with different  duration of data transmission ($\phi$).   }
\label{Figure2}
\end{figure}

\begin{figure}  
\centering
  \includegraphics[width=8.5cm,height=7.5cm]{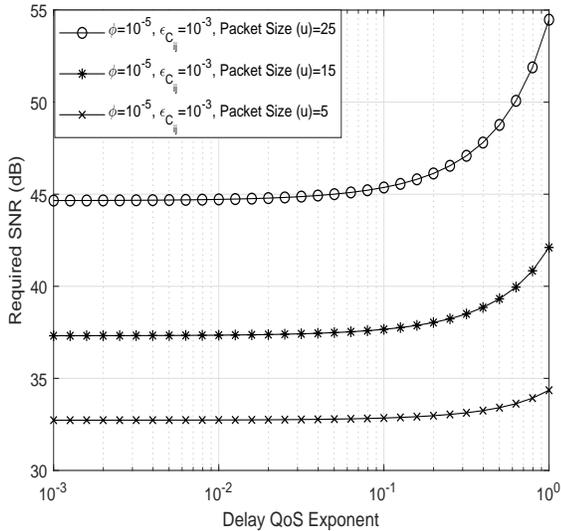}
\caption{SNR requirements for the given delay QoS exponent ($\theta_{ij}=10^{-3}$ to 1 )  with different  packet sizes.   }
\label{Figure3}
\end{figure}

\begin{figure} 
\centering
  \includegraphics[width=8.5cm,height=7.5cm]{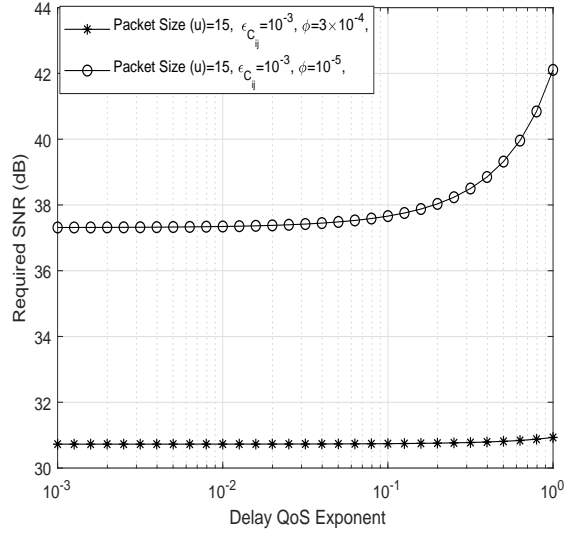}
\caption{SNR requirements for the given delay QoS exponent ($\theta_{ij}=10^{-3}$ to 1 ) with different  duration of data transmission ($\phi$).   }
\label{Figure4}
\end{figure}
In this section,  we analyse the effective-bandwidth of NOMA for URLLC.  Simulations have been performed to find the required SNR/SINR for a  certain error-rate and latency requirement. Simulation parameters have been described in Table \ref{tab_sim_para}, unless otherwise specified in different simulation results. 

For a given transmission-error probability and queueing-delay violation probability, the required SNR for the NOMA user has been analysed in detail. Queueing-delay with extremely short delay bound has been derived from the effective bandwidth using the Poisson process. In Fig. \ref{Figure1}, SNR requirements for the given transmission error probability ($\varepsilon_D= 10^{-5}$ to $10^{-3}$) and delay QoS exponent ($\theta_{ij}=0.1$) with differen packets sizes are shown. The required SNR is small with small packet size and as the packet size increases the required SNR increased. However, with increase in the channel errors (transmission error probability), the required SNR decreases. It shows that, for tighter reliability requirements, a higher SNR is needed.

 The duration of data transmission within one frame duration has also been analysed in Fig. \ref{Figure2}. The increase in data transmission duration with the  given frame duration of $T_{\rm{f}}=0.5 \rm{ms}$  also results into the lower requierd SNR. From Fig. \ref{Figure1} and \ref{Figure2}, it is clear that with the short packet and short data transmission duration the required SNR decreases.

 Queuing-delay for NOMA in URLLC with extremely short delay bound ($D_{\rm max}= 0.8 \rm{ms}$) has also been investigated through simulations.  In Fig. \ref{Figure3}, required SNR with the different values of delay QoS exponent has been shown. The increased value of $\theta_{ij}$ shows the increased in the delay (more stringent delay requirements). With increased in the delay, the required SNR increased. From $\theta_{ij}=0.1$ to $1$, the delay requirements are more stringent, which shows that the SNR requirements will also be high. This trend has also been investigated with different packet size of 5, 15, and 25 bytes while keeping the data transmission duration fixed. Here, also the short packets with different delay requirements show small  SNR requirements as compared to the large packets. The delay analysis for different data transmission duration with packet size of 15 bytes has been shown in Fig. \ref{Figure4}. The short data transmissions duration seems to be more susceptible with more stringent delay requirements as compared to the slighter larger data transmission duration. For this analysis the frame duration has been kept fixed at $T_{\rm{f}}=0.5 \rm{ms}$.

From Fig. \ref{Figure1} to \ref{Figure4}, the SNR requirements for the NOMA systems for ultra-reliable and low latency communications have been analysed while considering the different transmission error probability and queueing-delay violation probability. It is clear from the simulation results that employment of short packets to ensure the ultra-reliability and latency has a significant impact on SNR requirements to achieve the certain latency and error rate for NOMA in URLLC. 

As a future work, we will perform an-depth performance analysis of NOMA for URLLC communications with different arrival rate and by optimizing the different resources in the system. We will also integrate other state-of-the-art technologies such as cogntive radio networks \cite{R149,R152} and wireless sensor networks \cite{R151} with NOMA-URLLC to investigate the delay-sensitive applications. 

\section{Conclusion}
\label{sec:conclusion}
In this paper, the performance of the power-domain non-orthogonal multiple access (NOMA) for ultra-reliable and low latency communications (URLLC)  has been analysed with the help of effective-bandwidth. Queuing-delay and transmission delay has been considered to model the latency, while the reliability is modelled using the transmission error probability and queueing-delay violation probability. NOMA has significantly reduced the physical layer latency and improved the reliability for URLLC to support the time-critical applications. Extensive simulations have been carried out to show the SNR requirements while considering transmission error probability, delay QoS exponent, and different packet sizes.

\end{document}